\documentstyle[12pt,a4,epsf,rotate]{article}
\def\refjl#1#2#3#4#5#6{\bibitem{#1} #2, {\it #3} {\bf #4} (#5) #6.}
\def\refbk#1#2#3#4{\bibitem{#1} #2, {\it #3}, #4.}
\def\etal{{\it et al}}
\def\NP{Nucl. Phys.}

\def\PL{Phys. Lett.}
\def\PRL{Phys. Rev. Lett.}
\def\PR{Phys. Rev.}

\def\ZP{Z. Phys.}

\def\APNY{Ann. Phys., NY}
\def\NC{Nuovo Cimento}

\def\RPP{Rep. Prog. Phys.}

\def\PPNP{Prog. Part. Nucl. Phys.}

\newcommand{\eqn}[1]{(\ref{#1})}
\newcommand{\be}{\begin{equation}}
\newcommand{\ee}{\end{equation}}
\newcommand{\no}{\nonumber}
\newcommand{\bel}[1]{\be\label{#1}}
\newcommand{\ba}{\begin{array}{c}}
\newcommand{\bat}{\begin{array}{cc}}
\newcommand{\ea}{\end{array}}
\newcommand{\beqn}{\begin{eqnarray}}
\newcommand{\eeqn}{\end{eqnarray}}

\newcommand{\bi}{\begin{itemize}}
\newcommand{\ei}{\end{itemize}}
\newcommand{\rms}{\rm\scriptsize}

\catcode`\@=11
\def\@cite#1#2{\unskip\nobreak\relax
    \def\@tempa{$\m@th^{\hbox{\the\scriptfont0 #1}}$}%
    \futurelet\@tempc\@citexx}
\def\@citexx{\ifx.\@tempc\let\@tempd=\@citepunct\else
    \ifx,\@tempc\let\@tempd=\@citepunct\else
    \let\@tempd=\@tempa\fi\fi\@tempd}
\def\@citepunct{\@tempc\edef\@sf{\spacefactor=\the\spacefactor\relax}\@tempa
    \@sf\@gobble}

\def\citenum#1{{\def\@cite##1##2{##1}\cite{#1}}}
\def\citea#1{\@cite{#1}{}}

\newcount\@tempcntc
\def\@citex[#1]#2{\if@filesw\immediate\write\@auxout{\string\citation{#2}}\fi
  \@tempcnta\z@\@tempcntb\m@ne\def\@citea{}\@cite{\@for\@citeb:=#2\do
    {\@ifundefined
       {b@\@citeb}{\@citeo\@tempcntb\m@ne\@citea\def\@citea{,}{\bf ?}\@warning
       {Citation `\@citeb' on page \thepage \space undefined}}%
    {\setbox\z@\hbox{\global\@tempcntc0\csname b@\@citeb\endcsname\relax}%
     \ifnum\@tempcntc=\z@ \@citeo\@tempcntb\m@ne
       \@citea\def\@citea{,}\hbox{\csname b@\@citeb\endcsname}%
     \else
      \advance\@tempcntb\@ne
      \ifnum\@tempcntb=\@tempcntc
      \else\advance\@tempcntb\m@ne\@citeo
      \@tempcnta\@tempcntc\@tempcntb\@tempcntc\fi\fi}}\@citeo}{#1}}
\def\@citeo{\ifnum\@tempcnta>\@tempcntb\else\@citea\def\@citea{,}%
  \ifnum\@tempcnta=\@tempcntb\the\@tempcnta\else
   {\advance\@tempcnta\@ne\ifnum\@tempcnta=\@tempcntb \else \def\@citea{--}\fi
    \advance\@tempcnta\m@ne\the\@tempcnta\@citea\the\@tempcntb}\fi\fi}

\renewcommand{\thefootnote}{\fnsymbol{footnote}}
\begin{document}
\begin{titlepage}
\begin{flushright}
{FTUV/97-42}\\
{IFIC/97-42}\\
{hep-ph/9707347}\\
\end{flushright}
\vspace{2cm}
\begin{center}
{\large\bf Effective Field Theory Description\\ of the Pion Form 
Factor\footnote{Work supported
in part by CICYT, Spain, under Grant No. AEN-96/1718.}}\\
\vfill
{\bf Francisco Guerrero and Antonio Pich}\\[0.5cm]
 Departament de
 F\'{\i}sica Te\`orica, 
 IFIC, CSIC -- Universitat de Val\`encia \\
 Dr. Moliner 50,  E-46100 Burjassot (Val\`encia),
Spain\\[0.5cm]
\end{center}
\vfill
\begin{abstract}
Using our present knowledge on effective hadronic theories, short--distance
QCD information, the $1/N_C$ expansion, 
analyticity and unitarity, we derive an expression
for the pion form factor, in terms of $m_\pi$, $m_K$,
$M_\rho$ and $f_\pi$.
This parameter--free prediction provides
a surprisingly good description of
the experimental data up to energies of the order of 1 GeV.
\end{abstract}
\vspace*{1cm}
\vfill
\end{titlepage}

\setcounter{footnote}{0}
\renewcommand{\thefootnote}{\alph{footnote}}


\vspace{0.5cm}
\noindent {\bf 1. \, Introduction}
\vspace{0.5cm}

The QCD currents   
are a basic ingredient of the electromagnetic (vector)
and weak (vector, axial, scalar, pseudoscalar) interactions.
A good understanding of their associated hadronic matrix elements is then
required to control the important interplay of QCD in electroweak processes.
Given our poor knowledge of the QCD dynamics at low energies, one
needs to resort to experimental information, such as $e^+e^-\to$ hadrons \
or semileptonic decays.

At the inclusive level, the analysis of two--point function correlators
constructed from the T--ordered product of two currents, has been widely
used to get a link between the short--distance description in terms of quarks
and gluons and the hadronic long--distance world.
In this way, information on fundamental parameters, such as
quark masses, $\alpha_s$ or vacuum condensates, is extracted from 
the available phenomenological information on current matrix elements.
The lack of good experimental data in a given channel translates then
into unavoidable uncertainties on the obtained theoretical results.

At very low energies, the chiral symmetry constraints 
\cite{WE:79}
are powerful enough to
determine the hadronic matrix elements of the light quark currents
\cite{GL:84,GL:85,chpt:95,EC:95}. 
Unfortunately, these chiral low--energy theorems only apply to the
threshold region. 
At the resonance mass scale, the
effective Goldstone chiral theory becomes meaningless and the use of some
QCD--inspired model
to obtain theoretical predictions
seems to be unavoidable.

Obviously, to describe the resonance region, one needs to use an
effective theory with explicit resonance fields as degrees of freedom.
Although not so predictive as the standard chiral Lagrangian for
the pseudo-Goldstone mesons, the resonance chiral effective theory
\cite{EGPR:89} turns out to provide interesting results,
once some short--distance dynamical QCD constraints are taken into
account \cite{EGLPR:89}.

We would like to investigate how well the resonance region can be
understood, using our present knowledge on effective hadronic
theories, short--distance  QCD information and other important constraints 
such as  analyticity and
unitarity. Present experiments are providing a rich data sample
(specially on hadronic $\tau$ decays), which can be used to test  our
theoretical skills in this important, but poorly understood, region.

Our goal in this first letter is to study one of the simplest current
matrix elements, the pion form factor $F(s)$, defined 
(in the isospin limit) as
\be\label{eq:F_def}
\langle\pi^0\pi^-|\bar d \gamma^\mu u |\emptyset\rangle
\, = \,\sqrt{2}\, F(s) \, (p_{\pi^-}-p_{\pi^0})^\mu \, ,
\ee
with $s\equiv q^2 \equiv (p_{\pi^-}+p_{\pi^0})^2$.
At $s>0$, $F(s)$ is experimentally known from the decay
$\tau^-\to\nu_\tau\pi^-\pi^0$
and (through an isospin rotation) from $e^+e^-\to\pi^+\pi^-$, while the elastic 
$e^-\pi^+$ scattering provides information at $s<0$.

Theoretically, the pion form factor has been extensively investigated
for many years. Thus, many of our results are not new.
We want just to achieve a description of  $F(s)$ as simple as possible,
in order to gain some understanding which could be used in other more complicated
current matrix elements.

\vspace{0.5cm}
\noindent {\bf 2. \, Effective Lagrangian Results}
\vspace{0.5cm}

Near threshold, the pion form factor is well described by
chiral perturbation theory (ChPT). At one loop, it takes
the form \cite{GL:85}:
\bel{eq:chpt}
F(s)^{\mbox{\rms ChPT}}
= 1+\frac{2 L_9^r(\mu)}{f_\pi^2}\, s -\frac{s}{96\pi^2 f_\pi^2} 
\left[ A(m_\pi^2/s,m_\pi^2/\mu^2) + 
{1\over 2} A(m_K^2/s,m_K^2/\mu^2)  \right] \, ,
\ee
where the functions
\bel{loop_fun}
A(m_P^2/s,m_P^2/\mu^2) \, =\, 
\ln{\left( m^2_P/\mu^2\right)} + {8 m^2_P\over s} -
\frac{5}{3}  + \sigma_P^3 \,\ln{\left(\frac{\sigma_P+1}{\sigma_P-1}\right)}
\ee
contain the loop contributions, with the usual phase--space factor
\bel{eq:sigma}
\sigma_P \equiv \sqrt{1 - 4 m_P^2/s}\, ,
\ee
and $L_9^r(\mu)$ is an $O(p^4)$ chiral counterterm renormalized
at the scale $\mu$. 
The measured pion electromagnetic radius \cite{AM:86},
$\langle r^2\rangle^{\pi^\pm} = (0.439\pm 0.008)$ fm$^2$, fixes
$L_9^r(M_\rho)= (6.9\pm 0.7)\times 10^{-3}$.

A two loop calculation in the
$SU(2)\otimes SU(2)$ theory (i.e. no kaon loops) has been completed
recently \cite{GM:91,CFU:96}.

Using an effective chiral theory which explicitly includes 
the lightest octet of vector resonances, one can derive \cite{EGPR:89,EGLPR:89}
the leading effect induced by the $\rho$ resonance:
\bel{eq:VM_eff}
F(s)^{\mbox{\rms V}} = 1 +{F_V G_V\over f_\pi^2} {s\over M_\rho^2 -s} \, ,
\ee
where the couplings $F_V$ and $G_V$ characterize the strength of the
$\rho\gamma$ and $\rho\pi\pi$ couplings, respectively.
The resonance contribution appears first
at the next-to-leading order in the chiral expansion.
This tree--level result corresponds to the leading term in the
$1/N_C$ expansion, with $N_C=3$ the number of QCD colours.
Comparing it with Eq.~\eqn{eq:chpt}, it gives an explicit calculation
of $L_9$ in the $N_C\to\infty$ limit; chiral loops (and the associated scale
dependence) being suppressed by an additional $1/N_C$ factor.

Eq. \eqn{eq:VM_eff} was obtained imposing the constraint that 
the short--distance behaviour of QCD
allows at most one subtraction for the pion form factor
\cite{EGLPR:89}. However, all empirical evidence and theoretical prejudice
suggests that $F(s)$ vanishes sufficiently fast for $s\to\infty$
to obey an unsubtracted dispersion relation.
If this is the case, one gets the relation \cite{EGLPR:89}
(at leading order in $1/N_C$ and if higher--mass states are not considered)
$F_V G_V/f_\pi^2 = 1$,
which implies the well--known Vector Meson Dominance (VMD) expression:
\bel{eq:VMD}
F(s)^{\mbox{\rms VMD}} = {M_\rho^2\over M_\rho^2 -s} \, .
\ee

The resulting prediction for the $O(p^4)$ chiral coupling $L_9$,
\bel{eq:L_9}
L_9 = {F_V G_V \over 2 M_\rho^2} = {f_\pi^2 \over 2 M_\rho^2}
= 7.2 \times 10^{-3} \, ,
\ee
is in very good agreement with the phenomenologically extracted value.
This shows explicitly that the $\rho$ contribution is the dominant
physical effect in the pion form factor.

Combining Eqs. \eqn{eq:chpt} and \eqn{eq:VMD},
one gets an obvious improvement of the theoretical description of $F(s)$:
\bel{eq:VMD_loop}
F(s) = {M_\rho^2\over M_\rho^2 -s}
-\frac{s}{96\pi^2 f_\pi^2} 
\left[ A(m_\pi^2/s,m_\pi^2/M_\rho^2) + 
{1\over 2} A(m_K^2/s,m_K^2/M_\rho^2)  \right] 
\, .
\ee
The VMD formula provides the leading term in the $1/N_C$
expansion, which effectively sums an infinite number of local ChPT
contributions to all orders in momenta
(corresponding to the expansion of the $\rho$ propagator in powers of
$s/M_\rho^2$).
Assuming that $L^r_9(M_\rho)$ is indeed dominated by the $1/N_C$ result
\eqn{eq:L_9}, the loop contributions encoded in the functions
$A(m_P^2/s,m_P^2/M_\rho^2)$
give the next-to-leading corrections in a combined expansion in powers
of momenta and $1/N_C$.

\vspace{0.5cm}
\noindent {\bf 3. \, Unitarity Constraints}
\vspace{0.5cm}

The loop functions $A(m_P^2/s,m_P^2/M_\rho^2)$ contain 
the logarithmic corrections
induced by the final--state interaction of the two pseudoscalars.
The strong constraints imposed by analyticity and unitarity allow us to perform
a resummation of those contributions.

The pion form factor is an analytic function in the complex $s$ plane, except
for a cut along the positive real axis, starting at $s=4m_\pi^2$,
where its imaginary part develops a discontinuity.
For real values $s<4m_\pi^2$, $F(s)$ is real.
The imaginary part of $F(s)$, above threshold, corresponds to the
contribution of on--shell intermediate states:
\bel{eq:abs}
\mbox{\rm Im}F(s) =  \mbox{\rm Im}F(s)_{2\pi} + \mbox{\rm Im}F(s)_{4\pi}
  + \cdots + \mbox{\rm Im}F(s)_{K\bar K} + \cdots
\ee
In the elastic region ($s<16 m_\pi^2$), Watson final--state theorem 
\cite{WA:55} relates the imaginary part of $F(s)$ 
to the partial wave amplitude
$T^1_1$ for $\pi\pi$ scattering with angular momenta and isospin
equal to one:
\bel{eq:watson}      
\hbox{Im}F(s+i\epsilon) =
\sigma_\pi T^1_1 F(s)^\ast =
e^{i\delta^1_1} \sin{\delta^1_1} F(s)^\ast =
\sin{\delta^1_1} \left|F(s)\right| =
\tan{\delta^1_1} \,\hbox{Re}F(s)
\, .
\ee
Since $\hbox{Im}F(s)$ is real, the phase of the pion form factor
is the same as the phase $\delta^1_1$ of the partial wave amplitude $T^1_1$.
Thus, one can write a ($n$-subtracted) dispersion relation in the form:
\bel{eq:disp}
F(s) \, =\, \sum_{k=0}^{n-1} {s^k\over k!} {d^k\over ds^k}F(0)
+ \frac{s^n}{\pi} \int^{\infty}_{4m^2_\pi} \, \frac{dz}{z^n}
\, {\tan{\delta^1_1(z)} \, \hbox{Re}F(z)\over z-s-i\epsilon} \, ,
\ee
which has the well--known Omn\`es \cite{MU:53,OM:58} solution:
\bel{eq:omnes}
F(s) \, =\, Q_n(s)\,\exp{\left\{\frac{s^n}{\pi} \int^{\infty}_{4m^2_\pi} \, 
\frac{dz}{z^n} \, \frac{\delta^1_1(z)}{z-s-i\epsilon}\right\}} \, , 
\ee
where
\bel{eq:Q(S)}
\ln{Q_n(s)} \, =\,
\sum_{k=0}^{n-1} {s^k\over k!} \, {d^k\over ds^k}\ln{F}(0)
\, .
\ee
Strictly speaking, this equation is valid only below the inelastic 
threshold  ($s \leq 16 m_\pi^2)$. However, the contribution from the
higher--mass intermediate states is suppressed  by phase space.
The production of a larger number of meson pairs is also of
higher order in the chiral expansion.

Using the lowest--order ChPT result
\bel{eq:delta}
\sigma T^1_1=\frac{s \sigma_\pi^3}{96\pi f_\pi^2} \simeq \delta^1_1 \, ,
\ee
the integral in Eq. \eqn{eq:omnes} generates the one--loop function
$-s A(m_\pi^2/s,y)/(96\pi^2f_\pi^2)$, 
up to a polynomial [in $s/(4 m_\pi^2)$]
ambiguity, which depends on the number of subtractions applied\footnote{
Obviously, the finite integration does not give rise to any dependence
on $y\equiv m_\pi^2/\mu^2$.}.
Thus, the Omn\`es formula provides an exponentiation of the chiral logarithmic
corrections. 
The subtraction function $Q_n(s)$ (and the polynomial ambiguity)
can be partly determined
by matching the Omn\`es result to the one--loop ChPT result 
\eqn{eq:chpt}, which fixes the first two terms of its Taylor expansion;
it remains, however, a polynomial ambiguity at higher orders.
This means an indetermination order by order between the 
non-logarithmic part of the pion form 
factor which must be in $Q_n(s)$ and the one which must be in the 
exponential.

The ambiguity can be resolved to a large extent, by matching the
Omn\`es formula to the improved result in Eq.~\eqn{eq:VMD_loop},
which incorporates the effect of the $\rho$ propagator.
One gets then:
\bel{eq:naive_resum}
F(s)=\frac{M^2_{\rho}}{M^2_{\rho}-s} \,\exp{\left \{ \frac{-s}{96\pi^2 f^2} 
A(m_\pi^2/s,m_\pi^2/M_\rho^2) \right\}} \, .
\ee
This expression for the pion form factor satisfies all previous low--energy
constraints  and, moreover, has the right phase at one loop.

Eqs. \eqn{eq:naive_resum} has obvious shortcomings.
We have used an $O(p^2)$ approximation to the phase shift $\delta^1_1$,
which is a very poor (and even wrong) description in the higher side
of the integration region. Nevertheless, one can always take a  sufficient
number of subtractions to emphasize numerically the low--energy region.
Since our matching has fixed an infinite number of subtractions, the
result \eqn{eq:naive_resum} should give a good approximation for
values of $s$ not too large.
One could go further and use the $O(p^4)$ calculation of $\delta^1_1$
to correct this result. While this could (and should) be done, it would
only improve the low--energy behaviour, where Eq. \eqn{eq:naive_resum}
provides already a rather good description. However, we are more interested
in getting an extrapolation to be used at the resonance peak.

\vspace{0.5cm}
\noindent {\bf 4. \, The Rho Width}
\vspace{0.5cm}

The off-shell width of the $\rho$ meson can be easily 
calculated, using the resonance chiral effective theory
\cite{EGPR:89,EGLPR:89}. One gets:
\beqn\label{eq:rho_width}
\Gamma_\rho (s) &\! = &\! 
{M_\rho\, s\over 96\pi f_\pi^2} \, \left\{\theta (s-4m_\pi^2)\,\sigma_\pi^3
+{1\over 2} \theta (s-4m_K^2)\,\sigma_K^3\right\}
\no\\ &\! = &\!
-{M_\rho\, s\over 96\pi^2 f_\pi^2} \, 
\hbox{Im}\left[ A(m_\pi^2/s,m_\pi^2/M_\rho^2) + 
{1\over 2} A(m_K^2/s,m_K^2/M_\rho^2)  \right] \, .
\eeqn
%
At $s=M_\rho^2$,
$\Gamma_\rho (M_\rho^2) = 144$ MeV, which
provides a quite good approximation to the experimental
meson width, $\Gamma_\rho^{\mbox{\rms exp}} = (150.7\pm 1.2)$ MeV.

Eq. \eqn{eq:rho_width} shows that,
below the $K\bar K$ threshold, the $\rho$ width is proportional to the $O(p^2)$
$\pi\pi$ phase shift:
%
$\Gamma_\rho (s) = M_\rho \delta^1_1$.
%
Making a Dyson summation of the $\rho$ self-energy corrections amounts
to introduce the $\rho$ width into the denominator of the $\rho$
propagator, shifting the pole singularity to 
$\sqrt{s}=M_\rho - {i\over 2} \Gamma_\rho$.
However,
expanding the resulting propagator in powers of $s/M_\rho^2$,
it becomes obvious that
the one--loop imaginary correction generated by
the width contribution is exactly the same as the one already contained
in the Omn\`es exponential.
This suggests to shift the imaginary part of the loop function
$A(m_P^2/s,m_P^2/M_\rho^2)$ from the exponential to the propagator.
Including the small contribution from the intermediate state $K\bar K$,
one has then:
\beqn\label{eq:final}
F(s) &\!\!\! =&\!\!\! 
\frac{M^2_{\rho}}{M^2_\rho-s-iM_\rho \Gamma_\rho (s)} 
\exp \Biggl\{
\frac{-s}{96\pi^2 f_\pi^2} \biggl[\hbox{Re}A(m_\pi^2/s,m_\pi^2/M_\rho^2) 
\biggr.\Biggr.\no\\ &&\Biggl.\biggl.
\qquad\qquad\qquad\qquad\qquad\qquad\qquad
\hbox{} + \frac{1}{2}\hbox{Re} 
A(m_K^2/s,m_K^2/M_\rho^2)  \biggr] \Biggr\} \, .
\eeqn
This change does not modify the result at order $p^4$, which still 
coincides with ChPT, but makes the phase shift pass trough $\pi /2$ at the 
mass of the $\rho$ resonance.

Expanding Eq.~\eqn{eq:final} in powers of momenta, one can check
\cite{paco}
that it does a quite good job in generating the leading $O(p^6)$
contributions in the chiral expansion.
The known coefficients for linear and quadratic logarithms 
of the ChPT result \cite{CFU:96}, 
are well reproduced in the chiral limit.
While this is also true for Eq.~\eqn{eq:naive_resum},
the expansion of Eq.~\eqn{eq:final} gives a better description \cite{paco}
of the polynomial part of the $O(p^6)$ ChPT result.

\vspace{0.5cm}
\noindent {\bf 5. \, Numerical Results}
\vspace{0.5cm}

We can see graphically in figure 1 how Eq. \eqn{eq:final}
provides a very good description of the data
\cite{B:85} up to rather high energies.
The only parameters appearing in the pion form factor formula
are $m_\pi$, $m_K$, $M_\rho$ and $f_\pi$, which
have been set to their physical values. Thus, 
Eq.~\eqn{eq:final} is in fact a parameter--free prediction.
The extremely good agreement with the data is rather surprising.

At low energies the form factor is completely dominated by the
polynomial contribution generated by the $\rho$ propagator. Nevertheless,
the summation of chiral logarithms turn out to be crucial to get the
correct normalization at the $\rho$ peak. The exponential factor
in Eq.~\eqn{eq:final} produces a 17\% enhancement of
$|F(s)|$ at $s=M_\rho^2$.

\begin{figure}[bt]
\centering
\rotate[r]{\epsfysize=14cm\epsfbox{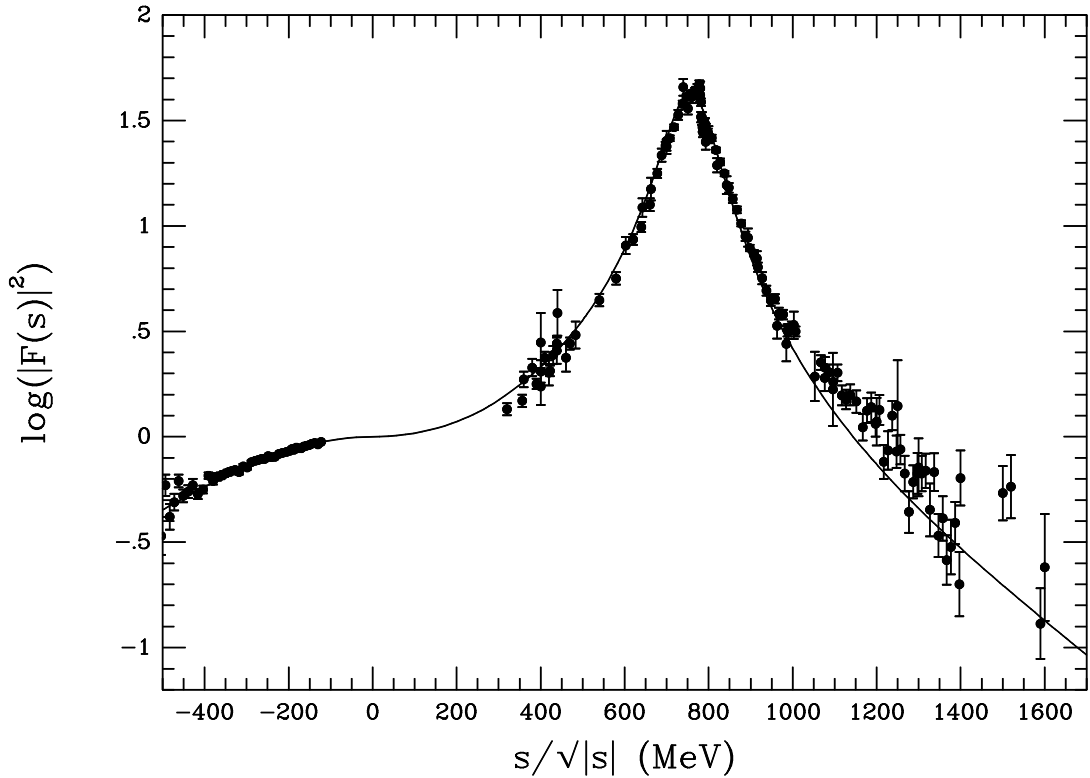}}
\caption{$|F(s)|^2$ (in logarithmic scale) versus $s/\sqrt{s}$.
The continuous curve shows the theoretical prediction
in Eq.~\protect\eqn{eq:final}.}
\end{figure}

The data at $s>0$ have been taken from $e^+ e^- \rightarrow \pi^+ \pi^-$; 
so in the vicinity of the $\rho$ peak 
there is a small isoescalar contamination due to the $\omega$
resonance. This contribution is well--known and generates a slight
distortion of the $\rho$ peak, which can easily be included
in the theoretical formula \cite{tesina}.
The effect 
cannot be appreciated
at the scale of the figure.

\begin{figure}[bt]
\centering
\rotate[r]{\epsfysize=14cm\epsfbox{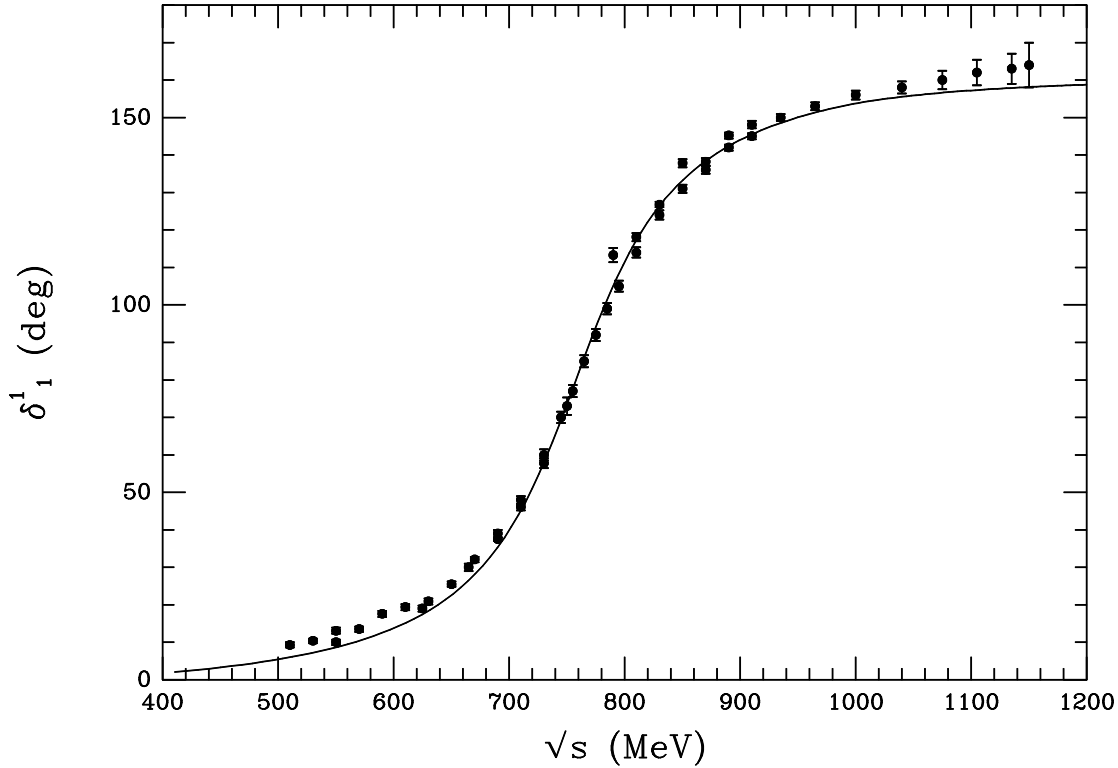}}
\caption{$\delta^1_1(s)$ versus $\sqrt{s}$. The
continuous curve shows the theoretical prediction
in Eq. \eqn{eq:final_delta}.}
\end{figure}

We also obtain a prediction for the phase shift $\delta^1_1$. From Eq. 
\eqn{eq:final} we get:
\bel{eq:final_delta}
\delta^1_1(s)\, =\, \arctan{\left\{\frac{M_{\rho} \Gamma_{\rho}(s)}{M^2_{\rho}-s}
\right\}} \, .
\ee
For $s<< M_\rho^2$, this expression reduces to the $O(p^2)$ ChPT result
in Eq.~\eqn{eq:delta}.
As shown in figure 2, the improvement obtained through Eq.~\eqn{eq:final_delta}
provides a quite good description of  
the experimental data \cite{P:73,E:74} over a rather wide energy range.
At large energies, the phase shift approaches the asymptotic limit
$\delta^1_1(s\to\infty) = \arctan{\{-\xi M_\rho^2/(96\pi f_\pi^2)\}}$.
If only the elastic $2\pi$ intermediate state is included, $\xi=1$ and
$\delta^1_1(s\to\infty) = 167^o$; taking into account
the $K\bar K$ contribution, one gets $\xi=3/2$, which slightly lowers the
asymptotic phase shift to $\delta^1_1(s\to\infty) = 161^o$.

\newpage
\vspace{0.5cm}
\noindent {\bf 5. \, Summary}
\vspace{0.5cm}

Using our present knowledge on effective hadronic theories, short--distance
QCD information, the $1/N_C$ expansion, 
analyticity and unitarity, we have derived a simple expression
for the pion form factor, in terms of 
$m_\pi$, $m_K$, $M_\rho$ and $f_\pi$.
The resulting parameter--free prediction
gives a surprisingly good description of
the experimental data up to energies of the order of 1 GeV.

Our main result, given in Eq.~\eqn{eq:final}, contains two
basic components. The $\rho$ propagator provides the leading
contribution in the limit of a large number of colours; 
it sums an infinite number of local terms in the low--energy
chiral expansion. Chiral
loop corrections, corresponding to the final--state 
interaction among the two pions, appear at the next order in the $1/N_C$
expansion; the Omn\`es exponential allows to perform a summation of
these unitarity corrections, extending the validity domain of the
original ChPT calculation.

Requiring consistency with the Dyson summation of the $\rho$ self-energy,
forces us to shift the imaginary phase from the exponential to the
$\rho$ propagator. While this change does not modify the rigorous
ChPT results at low energies, it does regulate the $\rho$ pole  and
makes the resulting phase shift pass through $\pi/2$ at the mass
of the $\rho$ resonance.

As shown in figures 1 and 2, the experimental pion form factor is
well reproduced, both in modulus and phase. Although the $\rho$
contribution is the dominant physical effect, the Omn\`es summation
of chiral logarithms turns out to be crucial to get the correct
normalization at the $\rho$ peak.

Many detailed studies of the pion form factor have been 
already performed previously
\cite{GM:91,CFU:96,GS:68,TR:88,KS:90,DT:96,HA:96,HA:97}.
The different ingredients we have used can in fact be found in the
existing literature on the subject. However, it is only when one combines
together all those physical informations that such a simple description of
$F(s)$ emerges.

Our approach can be extended in different ways (including two--loop ChPT
results, $\rho'$ contribution, \ldots). Moreover, one should investigate whether
it can be applied to other current matrix elements where the underlying physics
is more involved. 
We plan to study those questions in future publications.

\vspace{0.5cm}
\noindent {\bf Acknowledgements}
\vspace{0.5cm}

We would like to thank A. Santamar\'{\i}a for his help in the compilation
of the experimental data. 
Useful discussions with
 M. Jamin, J. Prades and E. de Rafael
 are also acknowledged.
 The work of
F. Guerrero has been supported by a FPI scholarship of
 the Spanish {\it Ministerio de Educaci\'on y Cultura}.

\end{document}